\documentclass[journal=apchd5, manuscript=article]{achemso}
\DeclareUnicodeCharacter{2212}{-}
\usepackage[T1]{fontenc}
\usepackage[utf8]{inputenc}
\usepackage{standalone}
\usepackage{textgreek}
\usepackage{transparent}
\usepackage{mathtools, nccmath}
\usepackage{tabularx,booktabs}
\usepackage{multicol}
\usepackage{tikz}
\usetikzlibrary{fit, matrix}
\usepackage{soul}
\usepackage{float}
\usepackage{lmodern}

\usepackage[version=3]{mhchem} 



\makeatletter
\let\l@addto@macro\relax
\makeatother
\usepackage[fontsize=9pt]{scrextend}
\AtBeginDocument{\singlespacing}

\author{Dmitriy~Yavorskiy}
\affiliation[1]{CENTERA Labs, Institute of High Pressure Physics, Polish Academy of Sciences, Sokołowska 29/37, 01-142 Warsaw, Poland}
\alsoaffiliation[4]{CENTERA, CEZAMAT, Warsaw University of Technology, Poleczki 19, 02-822 Warsaw, Poland}
\email{dmitriy.yavorskiy@unipress.waw.pl}

\author{Adil~Rehman}
\affiliation[1]{CENTERA Labs, Institute of High Pressure Physics, Polish Academy of Sciences, Sokołowska 29/37, 01-142 Warsaw, Poland}

\author{Wojciech~Brzezicki} 
\affiliation[5]{International Research Centre MagTop, Institute of Physics, Polish Academy of Sciences, Aleja Lotnikow 32/46, 02-668 Warsaw, Poland}
\alsoaffiliation[6]{Institute of Theoretical Physics, Jagiellonian University,
ulica S. Łojasiewicza 11, PL-30348 Kraków, Poland}

\author{Jan~Skolimowski} 
\affiliation[5]{International Research Centre MagTop, Institute of Physics, Polish Academy of Sciences, Aleja Lotnikow 32/46, 02-668 Warsaw, Poland}

\author{Marcin~Białek}
\affiliation[1]{CENTERA Labs, Institute of High Pressure Physics, Polish Academy of Sciences, Sokołowska 29/37, 01-142 Warsaw, Poland}

\author{Wojciech~Knap}
\affiliation[1]{CENTERA Labs, Institute of High Pressure Physics, Polish Academy of Sciences, Sokołowska 29/37, 01-142 Warsaw, Poland}
\alsoaffiliation[4]{CENTERA, CEZAMAT, Warsaw University of Technology, Poleczki 19, 02-822 Warsaw, Poland}

\author{Dawid Wutke} 
\affiliation[7]{National Synchrotron Radiation Centre SOLARIS, Jagiellonian University, Czerwone Maki 98, PL-30392 Cracow, Poland}

\author{Natalia Olszowska} 
\affiliation[7]{National Synchrotron Radiation Centre SOLARIS, Jagiellonian University, Czerwone Maki 98, PL-30392 Cracow, Poland}

\author{Andrzej~Wiśniewski} 
\affiliation[5]{International Research Centre MagTop, Institute of Physics, Polish Academy of Sciences, Aleja Lotnikow 32/46, 02-668 Warsaw, Poland}
\alsoaffiliation[2]{Institute of Physics, Polish Academy of Sciences, Aleja Lotników 32/46, 02-668 Warsaw, Poland}

\author{Ashutosh S. Wadge}
\affiliation[5]{International Research Centre MagTop, Institute of Physics, Polish Academy of Sciences, Aleja Lotnikow 32/46, 02-668 Warsaw, Poland}
\email{wadge@magtop.ifpan.edu.pl}

\title{Far-field terahertz spectroscopy across the charge-density-wave transition in 2H-NbSe$_2$}

\abbreviations{THz, CDW, NbSe$_2$}

\begin{document}

\begin{abstract}
Charge-density-wave (CDW) formation in 2H-NbSe$_2$ modifies the low-energy electronic structure and gives rise to collective excitations coupled to the lattice. Here, we investigate bulk 2H-NbSe$_2$ single crystals using far-field terahertz time-domain spectroscopy (THz-TDS) in reflection geometry across the CDW transition at $T_{\mathrm{CDW}} \approx 33$~K. Below $T_{\mathrm{CDW}}$, the THz response shows a pronounced high-frequency feature near 1.5~THz together with longer-lived sub-THz oscillations. Both responses progressively weaken upon warming and are strongly suppressed across the CDW transition, supporting their association with the CDW state. Using time-dependent Ginzburg--Landau simulations, we reproduce the main features of the experimental THz response, associating the high-frequency response mainly with CDW amplitude dynamics and the sub-THz response with defect-pinned phase dynamics. The $\sim1.5$~THz feature lies close to the frequency range reported for the Raman CDW amplitude mode, while coupling to lattice degrees of freedom may also influence its spectral position. We also performed complementary angle-resolved photoemission spectroscopy measurements, which reveal momentum-selective redistribution of near-Fermi-level spectral weight across the transition. Together, these results show that far-field THz spectroscopy provides a sensitive probe of collective CDW dynamics in bulk 2H-NbSe$_2$.

\end{abstract}

\section{Introduction}

Charge-density-wave (CDW) materials support collective excitations of the electronic order and the accompanying lattice distortion. These include amplitude modes, corresponding to oscillations of the CDW amplitude, and phase modes, corresponding to shifts of the CDW phase. Their dynamics can be strongly influenced by electron--phonon coupling, commensurability, screening, and disorder.\cite{Gruner1988,Littlewood1982,Monceau2012} Among layered transition-metal dichalcogenides, 2H-NbSe$_2$ is a prototypical CDW material with an incommensurate CDW transition at $T_{\mathrm{CDW}}\approx 33$~K and superconductivity emerging below about 7~K.\cite{Moncton1977, TYokoya2001, MZehetmayer2010} The CDW state in 2H-NbSe$_2$ has been extensively studied using diffraction, Raman spectroscopy, scanning tunnelling microscopy, and angle-resolved photoemission spectroscopy (ARPES).\cite{Moncton1977,Sooryakumar:1980,Measson2014,Grasset:2018,Arguello2014} ARPES measurements have revealed strongly momentum-dependent electron--phonon coupling and relatively subtle changes in the quasiparticle spectra across the CDW transition, showing that the electronic response to CDW formation varies considerably across momentum space.\cite{Valla2004} Raman spectroscopy has identified the CDW amplitude mode in the low-frequency range, typically around $35$--$50$~cm$^{-1}$ in bulk 2H-NbSe$_2$.\cite{JCTsangPRL1976,Sooryakumar:1980,Measson2014,Grasset:2018,Lin:2020} In addition to amplitude dynamics, the CDW phase can be strongly affected by defects and local pinning.\cite{EOhPRL2020}

In 2H-NbSe$_2$, spectroscopic studies have revealed coupling between CDW modes and lattice vibrations. Low-frequency Raman measurements showed a Fano-type interaction between an interlayer shear phonon and a CDW-related mode, indicating hybridization between lattice and CDW excitations.\cite{Mialitsin2011,Kumbhakar2026} A low-frequency overdamped oscillation has also been observed in time-resolved reflectivity and attributed to collective dynamics of CDW puddles.\cite{Kumbhakar2026} Nonlinear terahertz (THz) measurements further showed that CDW collective modes contribute to the third-harmonic response below $T_{\mathrm{CDW}}$.\cite{Feng2023} These results show that THz spectroscopy can also provide access to collective CDW dynamics in 2H-NbSe$_2$.

At the local scale, THz pump--probe scanning tunnelling microscopy has provided direct access to CDW phase dynamics in 2H-NbSe$_2$. Sheng \textit{et al.} observed CDW-related sub-THz oscillations over approximately $0.15$--$0.9$~THz, with frequencies and amplitudes that varied across the surface and were strongly influenced by individual atomic defects.\cite{Sheng:2024} Together with time-dependent Ginzburg-Landau modeling, these oscillations were interpreted as defect-pinned CDW phase excitations. This shows that the low-frequency CDW response can vary strongly on the local scale. In a far-field experiment, such local responses are averaged over a macroscopic area, so individual local modes may merge into a broader and less structured sub-THz response.

THz time-domain spectroscopy (THz-TDS) probes the low-frequency electrodynamic response and provides information complementary to Raman spectroscopy, as the two techniques couple differently with matter excitations. Previous THz studies of NbSe$_2$ have mainly focused on device applications and nonlinear THz dynamics,\cite{Li2022,Jiang:2023,Shein:2024,Feng2023} while temperature-dependent far-field THz spectroscopy of bulk 2H-NbSe$_2$ across the CDW transition has received much less attention.

Here, we investigate bulk 2H-NbSe$_2$ single crystals using THz-TDS in reflection geometry between 12 and 52~K, across the CDW transition at $T_{\mathrm{CDW}}\approx33$~K. Below the transition, the reference-subtracted reflected THz field shows a pronounced feature near 1.5~THz ($\sim50$~cm$^{-1}$), together with longer-lived sub-THz oscillations. Both responses weaken upon warming and are strongly suppressed across the CDW transition. Time-dependent Ginzburg-Landau simulations reproduce the main features of the experimental spectra. Within the model, the high-frequency response is associated mainly with CDW amplitude dynamics, while the sub-THz response is associated with defect-pinned phase excitations. The latter interpretation is consistent with recent atomic-scale THz-STM measurements of 2H-NbSe$_2$.\cite{Sheng:2024} Complementary ARPES measurements reveal a momentum-selective redistribution of spectral weight near the Fermi level across the transition. Together, these results show that far-field THz spectroscopy can probe collective CDW dynamics in bulk 2H-NbSe$_2$.

\section{Results and discussion}\label{sec2}
\subsection{Temperature-dependent far-field THz response across the CDW transition}

We investigate a bulk 2H-NbSe$_2$ single crystal with lateral dimensions of approximately $5 \times 3$~mm$^2$ and a thickness of approximately 300~$\mu$m using THz-TDS in reflection geometry. We collected time-domain reflection traces in the temperature range of 12 to 52~K, covering the CDW transition at $T_{\mathrm{CDW}}\approx 33$~K. More experimental details are in the Methods section. To highlight temperature-dependent changes, we subtract the reference time-domain trace measured at each temperature from the reference time-domain traces measured at 52~K (above $T_{\mathrm{CDW}}$)
\begin{equation}
\Delta E(t,T) = E(t,T) - E(t,52~\mathrm{K}).
\label{eq:deltaE}
\end{equation}

\begin{figure*}[ht!]
\centering
\includegraphics[width=\textwidth]{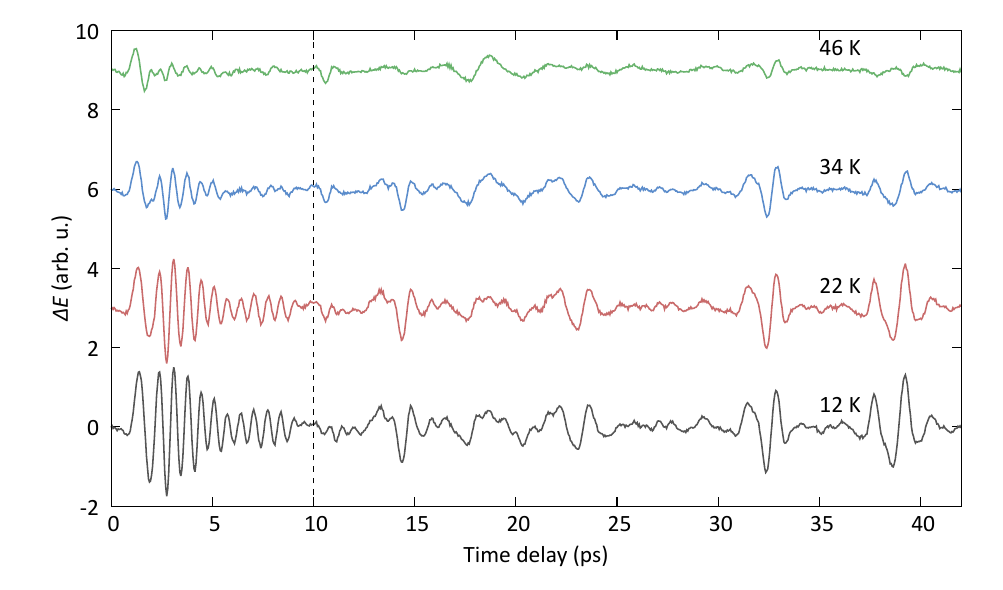}
\caption{
Temperature-dependent THz response of bulk 2H-NbSe$_2$. Reference-subtracted time-domain traces of the reflected THz electric field, $\Delta E(t,T)=E(t,T)-E(t,52~\mathrm{K})$, are shown at selected temperatures from 12 to 46~K. The time traces are vertically offset for clarity. At low temperatures, the response contains a fast oscillatory component at early time delays and longer-lived oscillations at later delays. Both components weaken
upon warming through the CDW transition at $T_{\mathrm{CDW}}\approx33$~K.
}
\label{fig1}
\end{figure*}

At 52~K, 2H-NbSe$_2$ is in the metallic normal state. Subtraction of the 52~K time-domain trace, measured in to the metallic normal-state response, suppresses the temperature-independent background and highlights the temperature-dependent response. Accordingly, $\Delta E(t,T)$ represents a differential reflected-field time-domain signal, and thus should not be interpreted directly as the absolute complex reflectivity or optical conductivity. The reference-subtracted time-domain traces at selected temperatures are shown in Fig.~\ref{fig1}. At low temperature, $\Delta E(t,T)$ contains a short-period oscillatory component at early time delays, with a period of approximately 0.7~ps. This component decays within the first $\sim10$~ps, and its amplitude decreases upon warming. At later delays, we observe additional oscillations with substantially longer periods. These late-time oscillations also become weaker with increasing temperature. The similar temperature evolution of the early- and late-time components, together with their pronounced enhancement below $T_{\mathrm{CDW}}$, suggests that both are associated with the CDW state, while their different time scales indicate distinct dynamical contributions.

\begin{figure}[ht!]
\centering
\includegraphics[width=\textwidth]{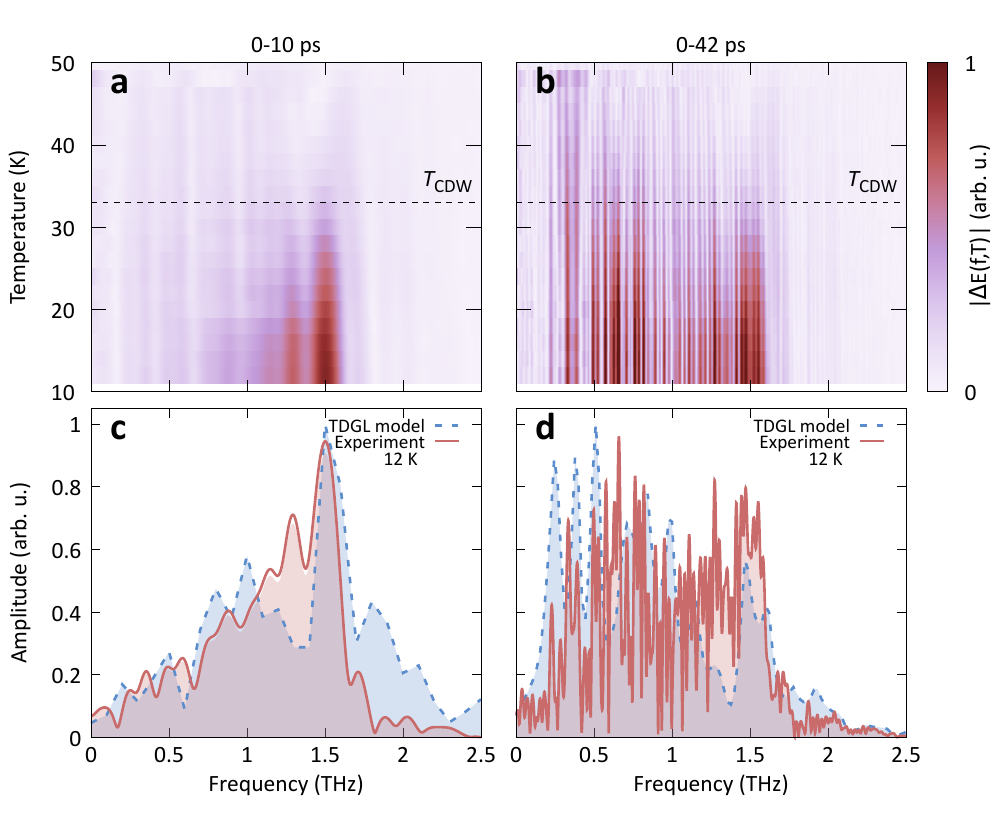}
\caption{
Time-window-dependent THz response and comparison with phenomenological TDGL simulations. 
(a) Temperature--frequency map of the Fourier amplitude obtained
from the $0$--$10$~ps interval of the reference-subtracted time traces. 
In (a) and (b), the vertical dashed line marks $T_{\mathrm{CDW}}\approx33$~K. 
(c,d) Comparison of the normalized experimental spectra at 12~K with the TDGL simulations for the (c) $0$--$10$~ps and (d) $0$--$42$~ps analysis windows, respectively.
}
\label{fig2}
\end{figure}

To examine the spectral content of the early-time response, we Fourier transform the traces within the $0$--$10$~ps interval. The corresponding spectra as a function of temperature are shown in Fig.~\ref{fig2}(a). At low temperature, we observe a structured high-frequency response consisting of several spectral features with a dominant peak near 1.5~THz. The overall response progressively weakens upon warming and is strongly reduced across the CDW transition. This temperature evolution supports associating the low-temperature THz response with CDW dynamics. 

The oscillations persisting at later time delays motivate analysis over an extended $0$--$42$~ps window. As shown in Fig.~\ref{fig2}(b), the extended analysis reveals a structured sub-THz response below approximately 1~THz, with several spectral maxima, together with the high-frequency response around 1.5~THz. The sub-THz response progressively weakens upon warming and is strongly suppressed across the CDW transition.

\subsection{Time-dependent Ginzburg-Landau modeling}
Here, we give the microscopic interpretation of the high- and low-frequency responses using phenomenological time-dependent Ginzburg-Landau (TDGL) modeling.
Following the approach of Ref.\ \cite{Sheng:2024}, with the amplitude mode coupled to a phase mode pinned by the impurities, we adopt a model with the total linear Lagrangian density given by
\begin{equation}
    {\cal L} = \sum_i\left\{
    \frac{1}{2}m_A\dot A_i^2 +\frac{1}{2}m_{\phi}\dot \phi_i^2 - aA_i^2-bA^4_i+c_A(A_{i+1}-A_i)^2\delta x^{-2}+c_{\phi}A_i^2(\phi_{i+1}-\phi_i)^2\delta x^{-2}+A_i\cos(iq_0\delta x+\phi_i)U_i
    \right\}
\end{equation}
where $A_i\cos(iq_0\delta x+\phi_i)$ is the charge density at site $i$ of a lattice (for simplicity assumed to be a one-dimensional chain along the density wave direction), $A_i$ is the amplitude of the CDW, $\phi_i$ is its phase, and $q_0$ its wavenumber. Taking time as measured in picoseconds and distance in nanometers, we adopt the parameters of the model from \cite{Sheng:2024} as:
$m_A=100$, $m_{\phi}=110$, $a=100(T-T_c)$, $b=0.2$, $c_A=2$, $c_{\phi}=0.11$, $q_0=5.9$ and $\delta x =0.1$. The impurity potential we take as $U_i=-U_0\exp{[-(i-i_0)^2/2]}$, where $i_0$ is the location of the (single) impurity and $U_0=125$. We simulate a system consisting of $100$ sites and average results over $20$ random positions of the impurity. For a complete description, we also add damping at the level of the equations of motion for $A_i$ and $\phi_i$ with damping coefficients $\gamma_A=240$ and $\gamma_{\phi}=40$ for amplitude and phase, respectively. We also add an external force of the form $F_{\phi}(t)=\kappa\sin(\Omega(t-t_0))\exp[(t-t_0)^2/\tau^2]$ that excites amplitude oscillations to model the THz pulse. Here we take: $\kappa = 0.14$, $\Omega=0.5$ and $\tau^2=10$.
The equations of motion for amplitude and phase then take the form $m_A\ddot{A_i}=\partial\mathcal{L}/\partial A_i - \gamma_A\dot{A}_i $ and
$m_\phi\ddot{\phi_i}=\partial\mathcal{L}/\partial \phi_i + F_{\phi}(t) - \gamma_\phi\dot{A}_i $.

We show the results of the TDGL simulation in Fig. \ref{fig2}(c-d). Here, we focus on the low-temperature case of $T=12$~K.
In the short $0\text{--}10\text{ ps}$ window, fast non-linear dynamics immediately following the THz drive, coupled with spectral Fourier broadening, result in a continuous, asymmetric feature centered around $1.2\text{--}1.5\text{ THz}$. Extending the analysis window to $0\text{--}42\text{ ps}$ sharpens the spectral resolution and allows early non-linear transients to decay, clearly separating the system's dynamics into two distinct regimes: a dominant high-frequency amplitude mode near $1.5\text{ THz}$ governed by the order parameter restoration force and a set of lower-frequency pinned phase modes around $0.3\text{--}0.5\text{ THz}$ sustained by defect pinning potentials. This time-window dependence demonstrates that while short-time THz probes capture the strongly driven, multi-mode electronic response, long-time windows reveal the intrinsic, weakly damped linear collective modes of the charge density wave condensate.

\subsection{Momentum-selective electronic reconstruction across the CDW transition}
\begin{figure}[ht!]
\centering
\includegraphics[width=1\textwidth]{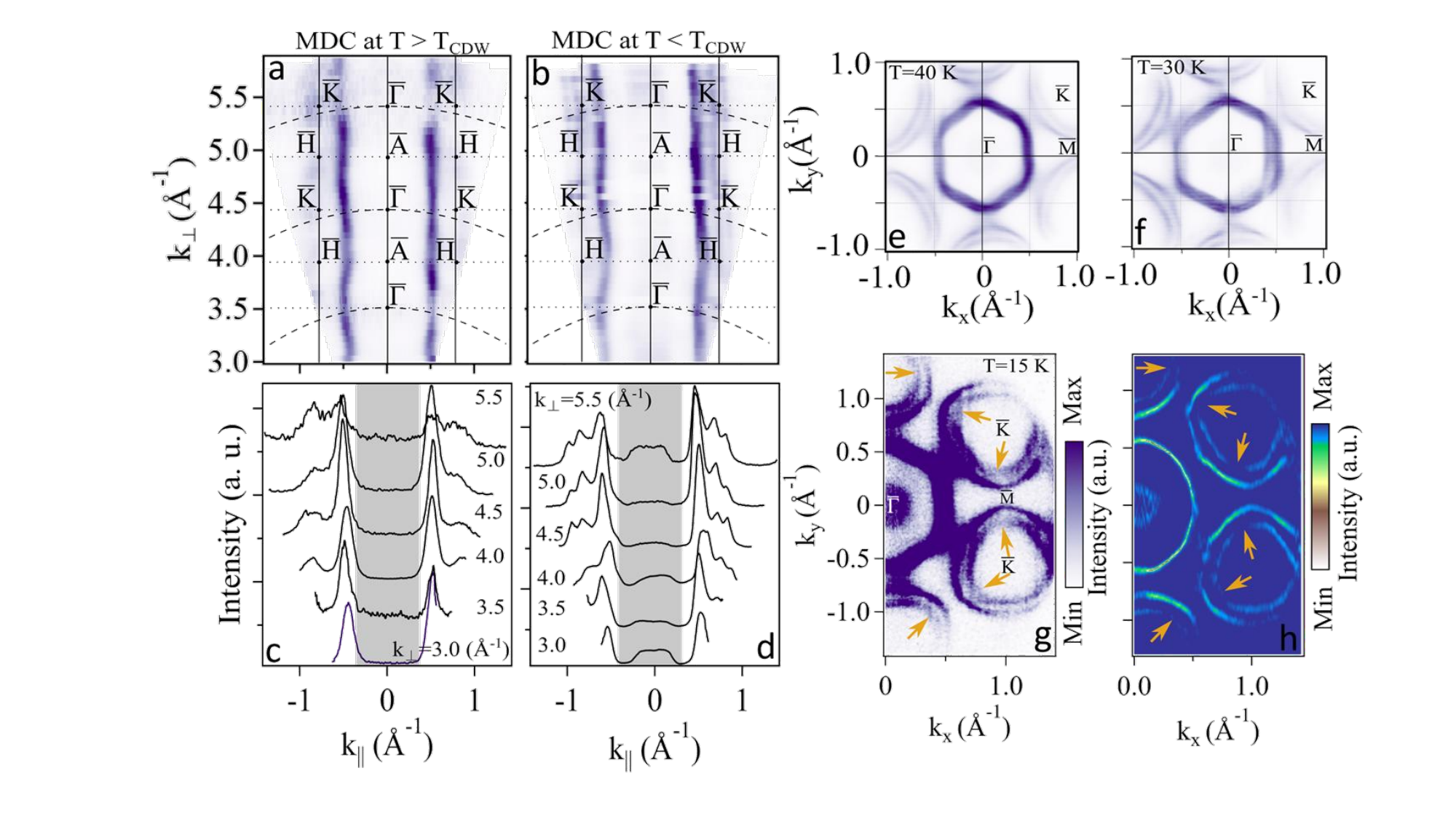}
\caption{
ARPES measurements of the momentum-dependent electronic structure of
2H-NbSe$_2$ across the CDW transition. (a,b) Momentum-resolved
near-Fermi-level intensity maps as a function of in-plane momentum
$k_{\parallel}$ and out-of-plane momentum $k_\perp$, measured above and below
$T_{\mathrm{CDW}}$, respectively. Dashed curves indicate Brillouin-zone
boundaries, and high-symmetry points are labeled. (c,d) Corresponding
momentum-distribution curves (MDCs) measured at selected values of
$k_\perp$, illustrating the evolution of spectral intensity and MDC peak
profiles. (e,f) Fermi-surface intensity maps measured at $T=40$~K and
$T=30$~K, respectively, using a photon energy of 64~eV. (g) Raw
low-temperature Fermi-surface intensity map measured at $T=15$~K using a
photon energy of 120~eV. Arrows indicate selected regions with
momentum-selective spectral-intensity modulations. (h) Two-dimensional
curvature analysis of the raw map in panel (g), which enhances the visibility
of intensity ridges and Fermi-surface contours. Arrows identify the same
momentum-space regions as in panel (g).
}
\label{fig3}
\end{figure}
To provide complementary momentum-resolved information on the electronic
structure across the CDW transition, we performed angle-resolved photoemission
spectroscopy (ARPES) measurements on crystals obtained from the same growth
batch as those used for the far-field THz experiments. Whereas THz reflection
spectroscopy probes the macroscopic low-energy electrodynamic response, ARPES
is a surface-sensitive and static probe of the momentum-resolved single-particle
electronic structure. The ARPES measurements are therefore used here to
characterize the electronic reconstruction accompanying CDW formation, rather
than to establish a direct microscopic assignment of the THz resonances.
Because reliable extraction of a well-defined CDW gap is not straightforward
under the present measurement conditions, we focus on the temperature-dependent
redistribution of spectral weight near the Fermi level.

\subsubsection{$k_\perp$-dependent electronic structure}

The three-dimensional character of the near-Fermi-level electronic structure
was examined by varying the photon energy and thereby probing different
out-of-plane momenta, denoted here as $k_\perp$. Figures~\ref{fig3}(a,b)
show momentum-resolved intensity maps as a function of the in-plane momentum
$k_{\parallel}$ and $k_\perp$, measured above and below
$T_{\mathrm{CDW}}$, respectively. The pronounced variation of the
near-Fermi-level spectral intensity with $k_\perp$ demonstrates that the
low-energy electronic structure has a substantial three-dimensional component.

Figures~\ref{fig3}(c,d) show momentum-distribution curves (MDCs) measured
at selected values of $k_\perp$ above and below $T_{\mathrm{CDW}}$,
respectively. The MDC peak profiles and spectral intensities evolve
systematically with $k_\perp$, confirming that measurements at a fixed photon
energy access specific out-of-plane momentum slices. This behavior is
consistent with previous photon-energy-dependent ARPES studies of bulk
2H-NbSe$_2$, which identified quasi-two-dimensional Nb $4d$-derived
Fermi-surface cylinders together with a more strongly $k_\perp$-dispersive,
pancake-shaped Se $4p_z$-derived pocket around the $\Gamma$ point.
\cite{Rossnagel2001} Photon-energy-dependent measurements are therefore
necessary for characterizing the three-dimensional electronic structure of
bulk 2H-NbSe$_2$.

\subsubsection{Temperature-dependent Fermi-surface evolution}

We next examine the temperature evolution of the Fermi-surface intensity maps.
Figures~\ref{fig3}(e,f) show maps measured at a photon energy of 64~eV at
$T=40$~K, above $T_{\mathrm{CDW}}$, and at $T=30$~K, below
$T_{\mathrm{CDW}}$, respectively. The overall Fermi-surface topology remains
broadly similar across the CDW transition. However, the distribution of
spectral intensity changes in localized momentum regions, indicating a
momentum-selective modification of the low-energy electronic structure.

At lower temperature, the raw Fermi-surface intensity map measured at
$T=15$~K using a photon energy of 120~eV
[Fig.~\ref{fig3}(g)] shows distinct intensity modulations along selected
segments of the Fermi contours, marked by arrows. Figure~\ref{fig3}(h)
shows the corresponding two-dimensional curvature analysis, which enhances the
visibility of the intensity ridges and Fermi-surface contours.
\cite{Zhang2011} The curvature representation is used as a visualization aid;
the momentum-dependent features discussed here are identified by comparison
with the corresponding raw intensity map in Fig.~\ref{fig3}(g).

The localized intensity changes are consistent with previously reported
momentum-selective CDW signatures in 2H-NbSe$_2$, including spectral-weight
suppression in restricted regions of the K-centered Fermi-surface barrels,
anisotropic gap effects, and subtle CDW-induced modifications of
Fermi-surface contours.\cite{Borisenko2009,Rahn2012,Kundu2024} The present
ARPES data thus provide complementary evidence that the CDW transition is
accompanied by a momentum-selective redistribution of near-Fermi-level spectral
weight within a three-dimensional electronic structure. These electronic
changes occur in the same temperature regime in which the far-field THz
measurements develop high-frequency and sub-THz CDW-associated responses.
Together, the two measurements provide a consistent, complementary picture of
the electronic reconstruction and collective electrodynamic changes associated
with CDW formation. However, they do not establish a mode-specific
correspondence between an individual ARPES feature and a particular THz
resonance.
\subsection{Discussion}


The temperature-dependent THz measurements reveal two distinct CDW-associated dynamical responses in bulk 2H-NbSe$_2$: a high-frequency feature near 1.5~THz and a longer-lived sub-THz response. The TDGL simulations suggest that the high-frequency feature is mainly related to CDW amplitude dynamics, whereas the lower-frequency response originates from phase dynamics affected by defect pinning. This provides a phenomenological interpretation of the two spectral components, which we discuss below in relation to previous spectroscopic studies.


The high-frequency feature is centered near 1.5~THz, corresponding to approximately 50~cm$^{-1}$. Raman studies of bulk 2H-NbSe$_2$ have reported the CDW amplitude mode around 40~cm$^{-1}$, with a broad amplitude-mode response extending towards 50~cm$^{-1}$ in some measurements.\cite{Sooryakumar:1980,Measson2014,Grasset:2018,Lin:2020} The frequency of the present THz feature is therefore close to, but somewhat higher than, the most commonly reported Raman amplitude-mode energy. We therefore do not assign the observed feature directly to the Raman amplitude mode. However, nonlinear THz third-harmonic-generation measurements have shown that CDW amplitude dynamics can contribute to the THz response below $T{\mathrm{CDW}}$.\cite{Feng2023} Together with the TDGL results, in which the high-frequency response is mainly associated with amplitude dynamics, this supports an amplitude-related interpretation of the observed resonance.


Low-frequency Raman studies have reported Fano-type coupling between an interlayer shear phonon and a CDW-related mode in 2H-NbSe$_2$, providing evidence that CDW collective dynamics can couple to lattice vibrations.\cite{Mialitsin2011,Kumbhakar2026} Strongly momentum-dependent electron--phonon coupling has also been observed by ARPES.\cite{Valla2004} These results make a contribution of lattice dynamics to the present $\sim1.5$~THz response plausible. However, our measurements do not resolve a specific phonon mode or direct signatures of mode hybridization. Together with the TDGL results, we therefore interpret the high-frequency feature primarily as an amplitude-related CDW response, while a possible contribution from lattice vibrations cannot be excluded.


The sub-THz response is consistent with defect-pinned CDW phase dynamics. Atomic-scale THz pump--probe scanning tunnelling microscopy has revealed CDW-related sub-THz excitations in 2H-NbSe$_2$ between approximately 0.15 and 0.9~THz, with frequencies and amplitudes that vary near individual atomic defects.\cite{Sheng:2024} Together with TDGL modeling, these excitations were interpreted as defect-pinned CDW phase modes. In the present far-field experiment, the response is averaged over a macroscopic area containing many defects and local CDW configurations. This spatial averaging can lead to a broad sub-THz response rather than a single well-defined resonance. The frequencies observed here therefore do not need to coincide directly with the local modes resolved by THz-STM.


Finally, the ARPES measurements show that the CDW transition is accompanied by a momentum-selective redistribution of near-Fermi-level spectral weight, while the overall Fermi-surface topology remains largely unchanged. This behavior is consistent with earlier ARPES studies reporting weak and momentum-selective CDW signatures in 2H-NbSe$_2$.\cite{Borisenko2009,Rahn2012,Kundu2024} The THz and ARPES measurements therefore provide complementary information on the collective response and the electronic reconstruction across the CDW transition. However, they do not establish a direct correspondence between a particular ARPES feature and an individual THz resonance.

\section{Conclusions}\label{sec3}

In this work, we investigated collective charge-density-wave dynamics in a
large-area 2H--NbSe$_2$ single crystal using terahertz time-domain spectroscopy
in reflection geometry. The CDW state exhibits distinct high-frequency and
longer-lived low-frequency collective responses, demonstrating that its
far-field electrodynamics cannot be described by a single uniform mode.

Time-dependent Ginzburg--Landau simulations reproduce the principal
time-window-dependent characteristics of the experimental spectra. Within the
TDGL framework, the high-frequency response is associated with the amplitude
sector of the CDW order parameter, whereas the lower-frequency response is
described by phase dynamics in the presence of impurity pinning. Complementary
angle-resolved photoemission spectroscopy measurements reveal a
momentum-selective redistribution of near-Fermi-level spectral weight across
the transition, providing electronic-structure context for the THz
observations.

Together, the experimental and modelling results establish broadband THz
reflection spectroscopy as a sensitive probe of collective CDW dynamics in
2H--NbSe$_2$. The measurements distinguish amplitude-related and
defect-sensitive phase-related responses in the CDW state, while further
experiments and microscopic modelling will be needed to determine the detailed
origin of the high-frequency collective excitation.

\section{Methods}

\subsection{Crystal growth}
Single crystals of NbSe$_2$ were prepared using the chemical vapor transport (CVT) technique. High-purity elemental niobium and selenium were weighed in stoichiometric proportions and loaded into an alumina crucible. The crucible was then enclosed in an evacuated quartz ampoule together with iodine (I$_2$) serving as the transport medium at a concentration of 3~mg/cm$^3$. Before sealing, the ampoule was flushed with ultra-high-purity argon gas to remove traces of oxygen and moisture, and subsequently evacuated to a pressure of approximately 10$^{-5}$~mbar to ensure an inert environment.

The sealed ampoule was placed in a three-zone horizontal tube furnace, where a temperature gradient was established between the source zone (850~$^\circ$C) and the growth zone (700~$^\circ$C). The system was maintained under these conditions for seven days to facilitate vapor transport and crystal growth. After the process, the furnace was allowed to cool naturally to room temperature.

The resulting product consisted of thick, shiny, hexagonal NbSe$_2$ flakes. Structural characterization using X-ray diffraction (XRD) confirmed the formation of single-phase 2H-NbSe$_2$ crystals as described in \cite{g8s4-dnxf}

\subsection{THz reflection spectroscopy}

THz reflection measurements were performed on a 2H--NbSe$_2$ single crystal with dimensions of approximately $3 \times 5 \times 0.3$~mm$^3$. The large and flat crystal surface was suitable for reflection measurements. The sample temperature was varied from 12 to 52~K in 1~K steps.

A broadband THz time-domain spectrometer (TOPTICA Photonics) equipped with photoconductive antennas was used to record the reflected THz electric field over a 50~ps time window. The incident THz beam was focused onto the NbSe$_2$ surface at an incidence angle of approximately $8^\circ$ using four 1-inch off-axis parabolic mirrors, which were also used to collect the reflected radiation. The THz beam path was continuously purged with dry nitrogen to reduce absorption by atmospheric water vapor. Frequency-domain spectra were obtained by applying a fast Fourier transform to the measured time-domain traces. The signals were zero-padded from 50 to 250~ps to provide a denser frequency grid in the Fourier-transformed spectra.

\subsection{ARPES measurements}
High-resolution angle-resolved photoemission spectroscopy (ARPES) measurements were carried out at the URANOS end station of the National Synchrotron Radiation Centre SOLARIS in Cracow, Poland. The beamline is equipped with a quasiperiodic, elliptically polarizing APPLE~II--type undulator that provides synchrotron radiation in the photon-energy range of 8--170~eV. Photoemitted electrons were detected using a SCIENTA OMICRON DA30L hemispherical electron spectrometer, offering an angular resolution of 0.1$^\circ$ and featuring deflectors that allow wide-angle band-structure mapping without the need to reposition the sample.

ARPES spectra were acquired using photon energies between 30 and 120~eV, covering the full Brillouin zone. All measurements were performed under ultrahigh-vacuum (UHV) conditions with a base pressure of about $3.7 \times 10^{-11}$~Torr. The data were collected at temperatures ranging from 286~K down to 20~K, ensuring stable thermal and spectral conditions throughout the experiment.

High-quality hexagonal bulk NbSe$_2$ flakes were used in this study. The crystals were mounted on pre-cleaned copper sample plates with conductive epoxy to ensure reliable electrical and thermal contact. The (001) surface of each crystal was initially exfoliated with Kapton tape under ambient conditions prior to insertion into the vacuum chamber. To minimize contamination during transfer, a fresh piece of Kapton tape was applied to the surface. After loading into the vacuum system, the tape was removed in the load lock, and the samples were exfoliated again \emph{in situ} under UHV conditions to expose a clean, atomically flat surface suitable for ARPES measurements.

\section*{Authors' contributions}
\begin{itemize}
    \item D.Y. contributed to the experiments, discussed the data interpretation, and contributed to writing the manuscript.
      
    \item A.R. conceived the idea, contributed to the experiments, and discussed the data interpretation.
    
    \item W.B. contributed to TDGL modeling, discussed the data interpretation, and contributed to writing the manuscript.

    \item J.K. contributed to TDGL modeling.
    
    \item M.B. contributed to the experiments.

    \item W.K. supplied resources.
    
    \item D.W. ARPES beam scientist

    \item N.O. ARPES beam scientist 
    
    \item A.W. supplied resources, ARPES experiment and discussion
    
     \item A.S.W. conceived the idea, grew the samples, discussed the data interpretation, ARPES experiment, and contributed to writing the manuscript.
\end{itemize}

\section*{Acknowledgements}
The research was supported by the European Union through the ERC-ADVANCED grant TERAPLASM (No. 101053716). Views and opinions expressed are, however, those of the author(s) only and do not necessarily reflect those of the European Union or the European Research Council Executive Agency. Neither the European Union nor the granting authority can be held responsible for them. We acknowledge the support of the "Center for Terahertz Research and Applications (CENTERA2)" project (FENG.02.01-IP.05-T004/23) and the “MagTop” Project (No. FENG.02.01-IP.05–0028/23) carried out within the "International Research Agendas" program of the Foundation for Polish Science co-financed by the European Union under the European Funds for a Smart Economy Programme. M. B. acknowledges the financial support of the Sonata No. BIS-13 2023/50/E/ST3/00584 grant of the National Science Centre of Poland. A.S.W acknowledges the support of the National Science Centre, Poland (NCN), through the MINIATURA 9 with project No. 2025/09/X/ST3/00809. A.S.W and A.W acknowledge research at the National Synchrotron Radiation Centre SOLARIS is supported by the Ministry of Science and Higher Education, Poland, under Contract No. 1/SOL/2021/2. 



\bibliography{bib}

\end{document}